\documentclass{optica-article}

\journal{opticajournal} 

\articletype{Research Article}

\usepackage{lineno}

\begin{document}

\title{Nonreciprocity in transmission mode with planar structures for arbitrarily polarized light}

\author{Michela F. Picardi,\authormark{1} Vera I. Moerbeek,\authormark{1} Mariano Pascale\authormark{1} and Georgia T. Papadakis\authormark{1,*}}

\address{\authormark{1}ICFO - The Institute of Photonics Sciences, Castelldefels, Barcelona, Spain}

\email{\authormark{*}georgia.papadakis@icfo.eu} 


\begin{abstract*} 
Approaching thermodynamic limits in light harvesting requires enabling nonreciprocal thermal emission. The majority of previously reported nonreciprocal thermal emitters operate in reflection mode, following original proposals by M. Green \cite{green_time-asymmetric_2012} and others. In these proposals, cascaded nonreciprocal junctions that re-direct each junction's emission towards a subsequent one are employed for efficient light-harvesting. Recently, simplified concepts have been proposed in solar photovoltaics \cite{park_reaching_2022} and thermophotovoltaics \cite{park_nonreciprocal_2022}, respectively, that leverage the concept of tandem junctions to approach thermodynamic limits. In these simplified scenarios, polarization-independent nonreciprocal response in transmission mode is required. We propose a pattern-free heterostructure that enables such functionality, using a magneto-optical material embedded between two dissimilar dielectric layers.
\end{abstract*}
\section{Introduction}
\par{Renewable energy devices that operate via light-harvesting have achieved impressive performance in recent years. Examples include solar photovoltaic systems (PVs) \cite{schygulla_wafer-bonded_2024}, that convert sunlight to electricity, and thermophotovoltaic systems (TPVs) \cite{lee_air-bridge_2022,lapotin_thermophotovoltaic_2022,lopez_thermophotovoltaic_2023,roux_main_2024,giteau_thermodynamic_2024}, that convert radiant heat to electricity. The conversion efficiency of solar PVs and TPVs is ultimately bounded by thermodynamics \cite{ries_complete_1983, de_vos_endoreversible_1993,giteau_thermodynamic_2023,greiner_thermodynamics_2012,shockley_detailed_1961, green_third_2006}. The maximum efficiency for solar energy harvesting, e.g. solar PV and solar TPV systems, is given by the Landsberg limit (93.3\%) \cite{landsberg_thermodynamic_1980}. By engineering the spectrum of emitted radiation by bodies other than the sun, TPV systems can approach the ultimate thermodynamic efficiency of a Carnot engine (95\% for solar irradiation) \cite{park_nonreciprocal_2022}.}

\par{Approaching these thermodynamic limits requires violating Kirchhoff's law of thermal radiation. Kirchhoff's law imposes a reciprocity constraint to the thermal emissivity of a material that ought to be equal to its absorptivity. The performance of standard reciprocal PVs and TPVs relies on maximizing absorption of photons at energies above the band gap of a PV junction. By enabling nonreciprocal emission, it becomes possible to also leverage the radiation emitted by the junction itself (arising from radiative recombination and incandescence). To approach the Landsberg limit, cascaded configurations of junctions have been proposed, for example in the seminal paper by M. Green \cite{green_time-asymmetric_2012}. In \cite{green_time-asymmetric_2012}, a configuration of nonreciprocal junctions is proposed such that each junction absorbs light at an angle of incidence and re-directs its own emission towards a subsequent junction at a different angle of emission. In the limit of infinite junctions, the Landsberg limit is reached.}

\par{Motivated by the promise of ultra-efficient light-harvesting, in recent years, nonreciprocity has been explored significantly both fundamentally, by identifying symmetries between emission and absorption \cite{guo_adjoint_2022, jalas_what_2013}, but also within the context of existing magneto-optical \cite{zhao_near-complete_2019,zhu_near-complete_2014,fernandes_enhancing_2023} and Weyl materials \cite{zhao_axion-field-enabled_2020,park_violating_2021,tsurimaki_large_2020,wu_nonreciprocal_2024,wang_maximal_2023,butler_broadband_2023,gu_near-unity_2024,li_polarization-independent_2024}. It is hence established that Kirchhoff's law can be violated per frequency and wavenumber. The first experimental demonstrations of nonreciprocal absorption \cite{liu_broadband_2023} and emission \cite{shayegan_direct_2023} has also been reported recently. Despite these advancements, however, all previous works on nonreciprocal emission remain limited to a particular angle of emission and absorption \cite{gu_near-unity_2024,wang_maximal_2023,butler_broadband_2023,li_polarization-independent_2024}. In addition, all aforementioned works report polarization-sensitive thermal emitters, operating for either transverse electric ($s$-) polarization \cite{wu_nonreciprocal_2024} or, most commonly, transverse magnetic ($p$-) polarization  \cite{zhao_near-complete_2019, shi_ultra-broadband_2024,zhao_axion-field-enabled_2020,wang_maximal_2023}, whereas maximizing light-harvesting efficiency requires nonreciprocal response for any polarization. Furthermore, the majority of previous proposals require considerable nanolithographic efforts as they rely on diffraction gratings \cite{zhao_near-complete_2019, zhao_axion-field-enabled_2020,tsurimaki_large_2020} and guided mode resonances in photonic crystals \cite{park_violating_2021,asadchy_sub-wavelength_2020}. Finally, with the exception of \cite{park_violating_2021}, all previously considered nonreciprocal thermal emitters for light-harvesting operate in reflection mode, absorbing and emitting light towards the same half-space, as initially envisioned in \cite{green_time-asymmetric_2012}.}
 
\par
\begin{figure}
\begin{center}
\includegraphics[width=0.2\textwidth]{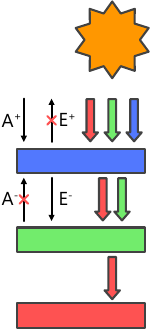}
  \caption{Schematic of nonreciprocal tandem junctions in transmission mode for high-efficiency photon-harvesting as proposed in \cite{park_reaching_2022 } and \cite{park_nonreciprocal_2022}, reaching the Landsberg and Carnot limit for solar PVs and TPV systems, respectively. The arrows denoted by $A^+$ and $E^-$ symbolise ideal absorption from the top and ideal emission to the bottom, respectively, whereas the arrows denoted by $A^-$ and $E^+$ symbolise suppressed absorption from the bottom and suppressed emission to the top, respectively. The energy bandgaps of the cells are decreasing from top to bottom, as indicated by colors, and each nonreciprocal cell directs its re-emitted radiation toward the cell underneath it.}
  \label{fig:fig1}
\end{center}
\end{figure}

{Recently, simplified configurations of series of junctions have been proposed, that do not require multiple reflections in order to reach thermodynamic limits. These mimic the concept of tandem solar cells, stacking PV junctions in a configuration that is more relevant in practice. As shown in \cite{park_reaching_2022} and \cite{park_nonreciprocal_2022}, a series of tandem nonreciprocal PV cells can reach the Landsberg and Carnot thermodynamic limit, respectively, provided that each cell absorbs radiation from one side (half-space) and emits towards the opposite side, referred to as operation in transmission mode henceforth. Other than \cite{park_reaching_2022,park_nonreciprocal_2022}, the same functionality is required for the absorber/emitter layer in the proposal by Jafari \textit{et al.}  \cite{jafari_ghalekohneh_nonreciprocal_2022}, which presents a simplified configuration for approaching the Landsberg limit with solar TPV systems. The requirement for optimal operation of a thermal emitter in transmission mode is schematically demonstrated in Fig. \ref{fig:fig1}; a cell maximally absorbs from the top ($A^+$ is maximum) while maximally emitting towards the bottom ($E^-$ is maximum). The only candidate for non-reciprocal semi-transparent thermal emission reported previously is discussed in \cite{park_violating_2021}, nonetheless the proposed nanostructure relies on a diffraction grating and is, thus, polarization-sensitive and lithographically complex, limiting the potential efficiency and practicality of potential devices. In addition, the device proposed in \cite{park_violating_2021} operates solely for normal incidence.} 

\par{Here, we introduce a pattern-free approach for violating Kirchhoff's law in transmission mode, for unpolarized light. The proposed thermal emitter/absorber is comprised of a magneto-optical material embedded between two dissimilar dielectric layers, and presents the same degree of nonreciprocal behavior for both linear polarizations, upon design optimization, for a broad range of angles of incidence/emission. The presented concept is general, hence the magneto-optical material can be either a conventional magneto-optical compound such as InAs or GaAs \cite{zhao_near-complete_2019,shayegan_direct_2023} or a Weyl semimetal as considered in various recent studies \cite{zhao_axion-field-enabled_2020,park_violating_2021,tsurimaki_large_2020,wu_nonreciprocal_2024,wang_maximal_2023,butler_broadband_2023,gu_near-unity_2024,li_polarization-independent_2024}. The operation of the device is tunable for both polarizations via in-plane rotation of the magneto-optical material or externally applied magnetic field with respect to the plane of incidence.}

\section{Results and Discussion}

\par
\begin{figure}
\begin{center}
\includegraphics[width=0.6\textwidth]{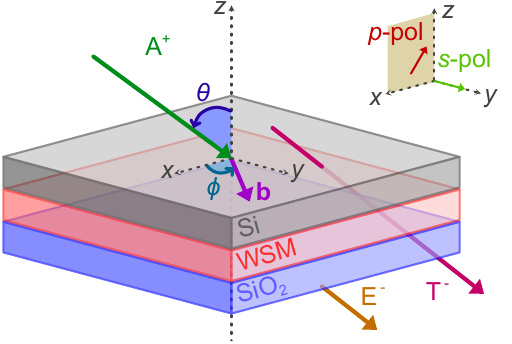}
  \caption{(a) Schematics of the proposed three-layer structure designed for nonreciprocal transmission, comprising a layer of magneto-optical material embedded between two dissimilar dielectrics. For the numerical results, we used the following materials: $Si$ layer with thickness $d_{Si} = 720 \text{nm}$, a layer of Weyl semimetal with thickness $d_{WSM} = 630 \text{nm}$, and a layer of $SiO_2$ with thickness $d_{SiO_{2}} = 425 \text{nm}$. The details on the dielectric permittivity for the WSM are provided in the Supplemental document.}
  \label{fig:schematics}
\end{center}
\end{figure}

{To break reciprocity between the thermal emissivity and absorptivity of a photonic structure in transmission mode, inversion symmetry ought to be broken about the propagation direction \cite{guo_adjoint_2022}, corresponding to $-z$ in Fig. \ref{fig:schematics}. Thereby, although even a single slab of nonreciprocal material, e.g. a magneto-optical material or Weyl semimetal, does break reciprocity in reflection mode \cite{wang_maximal_2023}, it does not suffice for breaking reciprocity in transmission mode. To break inversion symmetry with respect to the propagation direction, as discussed in \cite{asadchy_sub-wavelength_2020}, a planar heterostructure ought to be comprised of a minimum of three distinct layers. In \cite{asadchy_sub-wavelength_2020}, a photonic crystal composed of a three-layer unit cell was proposed for optical isolation. By contrast, here, we consider a simpler architecture of a structure made out of just three distinct layers. These layers consist of a magneto-optical material sandwiched in between two dielectric layers. This sequence of layers, as shown in Fig. \ref{fig:schematics}, was chosen to maximize index contrast between adjacent layers.}

\par{Without loss of generality, for the numerical simulations presented below, the dielectric layers are chosen to be silicon (Si) and silicon dioxide ($\mathrm{SiO}_2$) due to sufficiently different refractive indices at the wavelength of operation ($\lambda=10 \mu m$). These two layers have thicknesses of $d_{Si} = 720 \text{nm}$ and $d_{SiO_{2}} = 425 \text{nm}$, respectively. The magneto-optical material sandwiched in between the two dielectric layers is considered to be a Weyl semimetal and has a thickness of $d_{WSM} = 630 \text{nm}$, however we note that the proposed concept is general and applies to \textit{any} magneto-optical material. Weyl materials as well as magneto-optical materials exhibit a non-symmetric tensorial dielectric function containing conjugate pairs of off-diagonal tensor elements \cite{zhao_near-complete_2019,picardi_dynamic_2023,zhao_axion-field-enabled_2020}. Thereby, these materials break Lorentz reciprocity and can be used for nonreciprocal thermal emission \cite{butler_broadband_2023,tsurimaki_large_2020, zhao_near-complete_2019,shayegan_direct_2023}. In the case of conventional magneto-optical materials like InAs or GaAs, an externally applied magnetic field is required to induce the nonreciprocal behavior, however Weyl semimetals possess an intrinsic magnetic field arising from the separation between their Weyl nodes\cite{gorbar_electronic_2021,armitage_weyl_2018,han_giant_2022,konabe_anomalous_2024}, and thus require no external bias. The magnetic field, either externally applied or intrinsic, is indicated in Fig.\,\ref{fig:schematics} as $\mathbf{b}$. By aligning $\mathbf{b}$ with the $x$ axis, we describe the permittivity tensor $\overline{\overline{\varepsilon}}$ for a Weyl semimetal in Cartesian coordinates as:

\begin{equation}\label{eq:tensor_eps}
\overline{\overline{\varepsilon}} = \left(
    \begin{matrix}
       \varepsilon^d_{r} + i \varepsilon^d_i & 0 & + i \varepsilon^a\\
       0 & \varepsilon^d_{r} + i \varepsilon^d_i & 0\\
       - i \varepsilon^a & 0 &\varepsilon^d_{r} + i \varepsilon^d_i
    \end{matrix}\right),
\end{equation}
where $\varepsilon^d_{r}$, $\varepsilon^d_i$, and $\varepsilon^a \in \rm I\!R$. The details of the calculations of the dielectric function of the considered Weyl semimetal are provided in the Supplemental document, and the values for each of the parameters in Eq. \ref{eq:tensor_eps} are: $\varepsilon^d_{r}=-5.28$, $\varepsilon^d_i=$ 0.41, and $\varepsilon^a=14.79$, as in previous reports \cite{asadchy_sub-wavelength_2020}.}

\par
\begin{figure}
\begin{center}
\includegraphics[width=\textwidth]{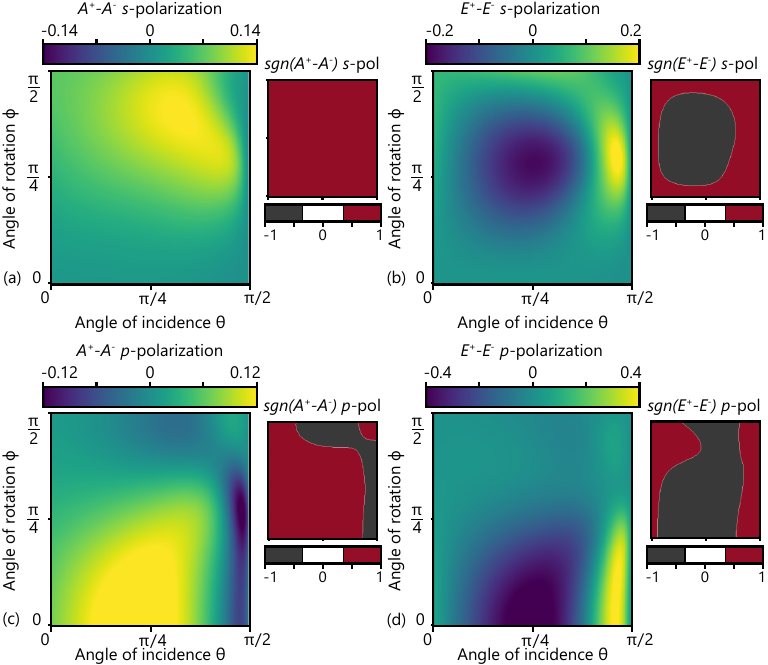}
  \caption{Differential absorption $A^+-A^-$ (a), (c) and emission $E^+-E^-$ (b),(d), between the two sides of the structure. The top row is for $s$-polarized light, while the bottom row is for $p$-polarized light. The insets show the sign of the difference, highlighting the large angular range for positive directional absorption and negative directional emission, required for solar energy conversion applications.} 
  \label{fig:diff_abs_em}
\end{center}
\end{figure}

{As already discussed in the introduction, and with respect to Fig. \ref{fig:fig1}, for light-harvesting applications, the targeted functionality of the considered structure is to achieve maximal values of $A^+$ and $E^-$, while conversely minimizing $A^-$ and $E^+$. Thereby, in Fig. \ref{fig:diff_abs_em}, we calculate the differential absorption and emission in the two half spaces (substrate and superstrate), defined as: $A^+(\theta,\phi,\rho) - A^-(\theta,\phi,\rho)$ and $E^+(\theta,\phi,\rho)- E^-(\theta,\phi,\rho)$, where $\theta$ is the angle of incidence in the $xz$ plane, $\phi$ is the in-plane rotation angle of the structure, and $\rho=s,p$ is the polarization. For light-harvesting as proposed in in \cite{park_reaching_2022,park_nonreciprocal_2022}, $A^+ - A^-$ ought to be positive, while $E^+ - E^-$ ought to be negative (Fig. \ref{fig:fig1}). In panels (a) and (c) of Fig. \ref{fig:diff_abs_em} is presented the quantity $A^+ - A^-$ for $s-$ and $p-$polarization, respectively, and in panels (b) and (d) the quantity $E^+ - E^-$ is presented for these polarizations. In the inset of each panel we explicitly report the sign of the respective quantity. As shown from panels (a)-(d) and their insets, for a broad range of incidence angles and in-plane rotation angles, $A^+-A^-$ remains positive for both polarizations while $E^+-E^-$ remains negative. As a consequence of operating in transmission mode, both the absorptivity $A^+$ as well as the emissivity $E^-$ have compromised values that considerably deviate from unity (Fig. \ref{fig:diff_abs_em}). This is expected as it is a direct consequence of energy balance \cite{park_violating_2021}. For maximal absorption from the upper half-space ($A^+$ in Fig. \ref{fig:schematics}) and maximal emission towards the lower half-space ($E^-$), energy balance equations dictate that maximal transmittance from the lower to the upper half space (indicated as $T^-$ in Fig. \ref{fig:schematics}) and minimal reflectance. In the Supplementary information, we present the values of the transmittance $T^-$. Indeed, in the angular range where $A^+$ and $E^-$ are maximal and opposite in sign, $T^-$ indeed deviates from zero.}

\begin{figure}
\begin{center}
\includegraphics[width=0.8\textwidth]{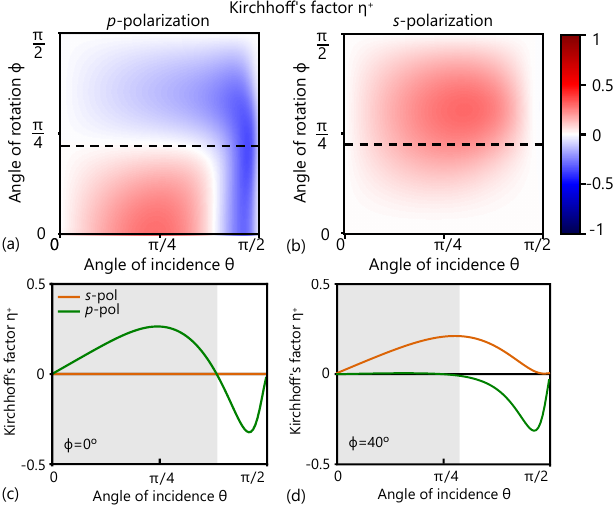}
  \caption{Kirchhoff's factor for $p$- and $s$-polarized light as a function of (a,b) angle of incidence and angle of rotation. Panels (c) and (d) are cross sections from (a) and (b) at specific angles. In (c), the Kirchhoff's factor is evaluated at $\phi=0$, where that the proposed structure is selectively nonreciprocal for $p$-polarized light only, for a broad range of angles of incidence, see shaded areas. Analogously, in (d), the nonreciprocal behaviour is suppressed for $p$-polarization, while the structure remains reciprocal for $s$-polarization for a broad range of angles of incidence. At these two angles the device operates in a polarization-selective mode.}
  \label{fig:results_figure}
\end{center}
\end{figure}

\par
\begin{figure}
\begin{center}
\includegraphics[width=0.5\textwidth]{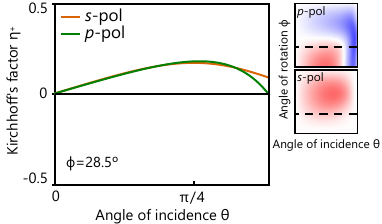}
  \caption{Kirchhoff's factor for $p$- and $s$-polarizations evaluated at $\phi=28.5^{\circ}$. At this in-plane rotation angle, the Kirchhoff's factor is equal in magnitude and sign for the two orthogonal polarizations. This operation mode is relevant for solar energy harvesting. Insets on the right side: same as panels a, b in Fig. \ref{fig:results_figure}.}
  \label{fig:equal_KF}
\end{center}
\end{figure}

\begin{figure}
\begin{center}
\includegraphics[width=0.5\textwidth]{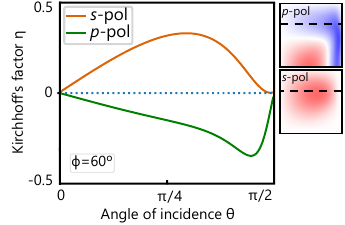}
  \caption{Kirchhoff's factor for $p$- and $s$-polarizations evaluated at $\phi=60^{\circ}$. For this in-plane rotation angle the Kirchhoff's factor is similar in magnitude but opposite sign for the two orthogonal polarizations, constituting the polarization-sensitive optical circulator mode. Insets on the right side: same as panels a, b in Fig. \ref{fig:results_figure}.}
  \label{fig:circulator_mode}
\end{center}
\end{figure}

{The metric commonly used to quantify nonreciprocity is the Kirchhoff's factor, $\eta$, which is defined as the difference between absorption, $A(\theta,\phi,\rho)$, and emission  $E(\theta,\phi,\rho)$, occurring in the same half-space (denoted as $+$ or $-$):
\begin{equation}
    \eta^{\pm}=A^{\pm}(\theta,\phi,\rho) - E^{\pm}(\theta,\phi,\rho).
\end{equation}
While the magnitude of $\eta$ quantifies the degree of nonreciprocal behavior, its sign determines the direction of the photon flow \cite{jafari_ghalekohneh_controlling_2024}. A value of $\eta=0$ indicates a reciprocal emitter, where absorption and emission are equal. A negative value of $\eta$ indicates that the structure is preferentially emitting, and, conversely, a positive value of $\eta$ shows that the structure absorbs more than it emits, as is required for light-harvesting in tandem configurations as discussed above (Figs. \ref{fig:fig1}, \ref{fig:diff_abs_em}, relative to \cite{park_reaching_2022,park_nonreciprocal_2022}). In Figs. \ref{fig:results_figure}, \ref{fig:equal_KF}, \ref{fig:circulator_mode}, we present values of $\eta$ for a broad range of angles of incidence, $\theta$, and in-plane rotations of the crystal axes of the Weyl semimetal, $\phi$, and discuss four different modes of operation.}

\par{In Fig. \ref{fig:results_figure}, we present $\eta^{+}$, the Kirchhoff factor for incidence from the upper half-space ($\theta$ ranges from $0$ to $\pi/2$ with respect to Fig. \ref{fig:schematics}) for $p-$, panel (a), and $s-$, panel (b), polarizations, respectively, for the entire range of in-plane rotation angles $\phi$. In panels (c) and (d), we show the operation of the device as selectively nonreciprocal for $p$-polarization, at $\phi=0$ and $s$-polarization at $\phi=40^{\circ}$, respectively (see shaded areas). First, for $\theta=0$ and $\theta=\pi/2$, $\eta=0$ for both $p-$, panel (a), and $s-$, panel (b), polarization and for all in-plane rotation angles ($\phi$), as inversion symmetry is not broken for these incidences. Indeed, from adjoint Kirchhoff's law \cite{guo_adjoint_2022}, the relationship $\eta(\theta) = -\eta(-\theta)$ holds, leaving $\eta=0$ the only possible solution for a planar structure. At $\phi=0$ for oblique incidence, nonreciprocity is only enabled for $p-$ polarization; this is expected since the electric field for $p-$polarization lies in the $xz$ plane and thus couples to the off-diagonal tensor elements of the Weyl semimetal, $\epsilon_{xz}$, $\epsilon_{zx}$ (Eq. \ref{eq:tensor_eps}). With respect to the coordinate system in Fig. \ref{fig:schematics}, the electric field for $s-$polarized light contains only a $y-$component, thus, in the absence of a rotation of the Weyl semimetal about the $z-$axis, $s$-polarized light does not experience its bi-anisotropy. This is demonstrated in panel c for $\phi=0$. By contrast, as shown in panel d, for $\phi=40^{\circ}$, $\eta$ is maximal for $s-$polarization while vanishing for $p-$polarization. Once the Weyl semimetal is rotated about its optical axis, the four diagonal tensor elements in the permittivity tensor that were initially null (Eq. \ref{eq:tensor_eps}) acquire nonzero values thus enabling nonreciprocity for $s$-polarization. As shown in Fig. \ref{fig:equal_KF}, upon appropriate selection of the rotation angle $\phi$, both polarizations exhibit the same degree of nonreciprocity, i.e. $\eta_{s}=\eta_{p}$, simultaneously.}

\par{In particular, Fig. \ref{fig:equal_KF} pertains to $\phi=28.5^{\circ}$. As shown, at this angle of rotation, which corresponds physically to the orientation of the intrinsic magnetic field of the Weyl semimetal ($\mathbf{b}$ in Fig. \ref{fig:schematics}) with respect to the coordinate system, for a large range of angles of incidence from $\theta=0$ to $\theta=60$, the Kirchhoff factor remains positive and equal for both linear polarizations. The Kirchhoff factor reaches a maximum of 0.19 when $\theta \approx 45^{\circ}$. This response is required for nonreciprocal tandem solar PVs or TPV systems as discussed in \cite{park_reaching_2022} and \cite{park_nonreciprocal_2022}, as explained in the introduction.}

\par{Other than a positive Kirchhoff factor for nonreciprocal emitters in transmission mode, as shown in Fig. \ref{fig:results_figure}, for angles of in-plane rotation between $40^{\circ}$ and $90^{\circ}$, the Kirchhoff factor for $p-$ polarization is of opposite sign to that for $s-$polarization. In this mode of operation, the structure can perform as a polarization-controlled optical circulator, since information encoded into photon flux of opposite linear polarizations will flow towards opposite preferential directions.  Specifically, $s$-polarized light is preferentially absorbed from the top and emitted from the bottom, viceversa $p$-polarized light is preferentially absorbed from the bottom and emitted from the top. We show an example of this behaviour in Fig. \ref{fig:circulator_mode}, where the Kirchhoff's factor is presented for $\phi=60^{\circ}$.}

\par{To summarize, we presented a three-layer heterostructure that breaks Kirchhoff's law of thermal radiation in transmission mode, for both linear polarizations for a broad range of angles of incidence and emission. Other than this functionality, which is relevant for light-harvesting at thermodynamic limits \cite{park_nonreciprocal_2022, park_reaching_2022}, the proposed devices can act as preferential thermal emitters for either linear polarization or as polarization-controlled optical circulators. The devices' parameter that defines the mode of operation is the in-plane rotation of the crystal axis of the Weyl semimetal with respect to the coordinate system. In the case of a magneto-optical material, this can be externally controlled via the application of a magnetic field. The structural simplicity of the proposed devices with respect to previous considerations make them relevant for proof-of-principle experimental demonstrations of nonreciprocity in thermal radiation, in transmission mode. In addition, owning to the structural simplicity of the proposed design, numerous optimised structures could be developed, exploiting, for example, Fabry-Perot resonances of the structure, or extending the design to layered photonic crystals, for light-harvesting as well as information-control applications.}

\begin{backmatter}
\bmsection{Funding}
MFP Acknowledges support from the Optica Foundation 20th Anniversary Challenge Award. MFP and GTP received the support of fellowships from “la Caixa” Foundation (ID 100010434). The fellowship codes are LCF/BQ/PI23/11970026 and LCF/BQ/PI21/11830019.
GTP also acknowledges support from the Spanish MICINN (PID2021-125441OA-I00, PID2020-112625GB-I00, and CEX2019-000910-S), Generalitat de Catalunya (2021 SGR 01443), Fundació Cellex, and Fundació Mir-Puig. 

\bmsection{Acknowledgements}
VIM acknowledges Dr Lu Wang for fruitful conversations. 

\bmsection{Disclosures}
The authors declare no conflicts of interest.

\bmsection{Data availability} No data were generated or analyzed in the presented research.




\end{backmatter}

\bibliography{sample}
\end{document}


\maketitle

\section{Permittivity of WSM}

\begin{figure}[htbp]
\centering
\fbox{\includegraphics[width=.9\linewidth]{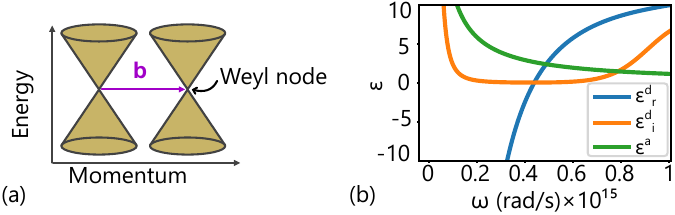}}
\caption{(a) Schematics of the band structure of a Weyl semimetal. (b) Dispersion of the permittivities of the Weyl semimetal, computed using Eqs. \ref{eq:eps_a} and \ref{eq:eps_d} using parameters from \cite{asadchy_sub-wavelength_2020}.}
\label{fig:supp}
\end{figure}

The intrinsic magnetic field originates from the anomalous Hall effect induced by the Weyl nodes separation in momentum $\mathbf{b}$ \cite{gorbar_electronic_2021,armitage_weyl_2018,han_giant_2022,konabe_anomalous_2024}, depicted schematically in Figure \ref{fig:supp}

In these materials, the displacement field $\mathbf{D}$ can be written as \cite{guo_light_2023}:
\begin{equation}\label{eq:displacement_field}
    \mathbf{D} = \varepsilon(\omega) \mathbf{E} - 2 \alpha (\mathbf{b}\times \mathbf{E}) = \overline{\overline{\varepsilon}}\,\mathbf{E},
\end{equation}
where $\mathbf{E}$ is the electric field and $\alpha = \frac{e^2}{2 \pi \hbar \omega}$, $e$ being the unitary charge. From Eq. \ref{eq:displacement_field} we see that the permittivity tensor of these materials contains off-diagonal terms, which depend on the static intrinsic magnetic field and can be written as:
\begin{equation}\label{eq:eps_a}
    \varepsilon^a = \frac{2 b e^2}{\pi \hbar \omega}.
\end{equation}
This static magnetic field suffices to obtain strong magneto-optical effects at optical frequencies, \cite{guo_light_2023} without the need of applying an external magnetic field, making these materials desirable.
The diagonal tensor elements can be calculated using the Kubo-Greenwood formula, given by:
\begin{equation}\label{eq:eps_d}
    \varepsilon^d = \varepsilon_0 \varepsilon_b + \frac{i r_s g \varepsilon_0}{6 Re(\Omega)}\Omega G\left(\frac{E_F \Omega}{2}\right) - \frac{r_s g \varepsilon_0}{6 \pi Re(\Omega)} \left[ \frac{4}{\Omega} + \frac{\pi^2}{3} \left(\frac{k_B T}{E_F(T)}\right) + 8\Omega \int_0^{\xi_c}\frac{G(E_F \xi) - G\left(\frac{1}{2}E_F \Omega\right)}{\Omega^2 - 4 \xi^2}\xi d\xi \right],
\end{equation}
where $\varepsilon_b$ is the background permittivity, $r_s = \frac{e^2}{2 \pi \varepsilon_0 \hbar v_F}$ is the fine structure constant, with $v_F$ being the Fermi velocity, $g$ is the number of Weyl nodes, $\Omega = \frac{\hbar(\omega + i \tau^{-1})}{E_F}$ is the normalised complex frequency with $E_F$ being the chemical potential and $\tau^{-1}$ the damping rate, $k_B$ the Boltzmann constant, $G(E) = n(-E)-n(E)$, where $n(E)$ is the Fermi distribution function and $\xi_c$ is the cutoff energy beyond which the band dispersion is no longer linear. 

\section{Transmission}

\begin{figure}[h!tbp]
\centering
\fbox{\includegraphics[width=0.9\linewidth]{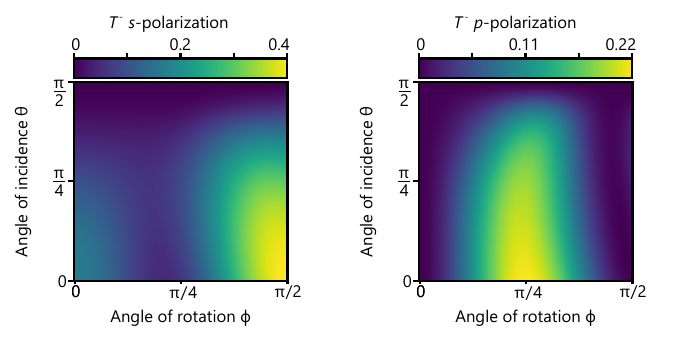}}
\caption{Transmission $T^-$ for $s$- (left) and $p$-polarized light (right).}
\label{fig:supp2}
\end{figure}

\bibliography{sample}
